\documentclass[11pt]{article}

\usepackage{fullpage,calc,amsfonts,amssymb,amsmath,bm,url,color,theorem}
\usepackage{graphicx,subfigure}
\usepackage{epsfig,multirow}
\usepackage{algorithmic,algorithm}
\usepackage{setspace}
\usepackage{epstopdf}

{\begin{list}{}{
    \settowidth{\labelwidth}{\mbox{\textnormal{#1}}}%
    \setlength{\leftmargin}{\labelwidth+\labelsep}}}%
{\end{list}}

    \makeatletter
    \def\multilimits@{\bgroup
  \Let@
  \restore@math@cr
  \default@tag
 \baselineskip\fontdimen10 \scriptfont\tw@
 \advance\baselineskip\fontdimen12 \scriptfont\tw@
 \lineskip\thr@@\fontdimen8 \scriptfont\thr@@
 \lineskiplimit\lineskip
 \vbox\bgroup\ialign\bgroup\hfil$\m@th\scriptstyle{##}$\hfil\crcr}
    \def\Sb{_\multilimits@}
    \def\endSb{\crcr\egroup\egroup\egroup}
\makeatother

\theorembodyfont{\rmfamily}

\begin{document}

\bibliographystyle{IEEEtran}
\vspace*{0.25in}

\begin{center}
\doublespacing
{\LARGE
Some Proof Derivations and Further Simulation Results for
``Semidefinite Relaxation and Approximation Analysis of a Beamformed Alamouti Scheme for Relay Beamforming Networks''} \\
~ \\
\singlespacing
Technical Report \\
Department of Systems Engineering and Engineering Management, \\
The Chinese University of Hong Kong,
Hong Kong \\
~ \\
Sissi Xiaoxiao Wu$^\S$,  Anthony Man-Cho So$^\S$, Jiaxian Pan$^\dag$, and Wing-Kin Ma$^\dag$ \\
~ \\
\today
~ \\ ~ \\
{\footnotesize
\begin{tabular}[t]{l}
$^\S$Department of Systems Engineering and Engineering Management, The Chinese University of Hong Kong, \\
Shatin, Hong Kong. \\
E-mail: \{xxwu, manchoso\}@se.cuhk.edu.hk.
\\
$^\dag$Department of Electronic Engineering, The Chinese University of Hong Kong, Shatin, Hong Kong. \\
E-mail: jxpan@ee.cuhk.edu.hk, wkma@ieee.org. \\
\end{tabular}}
\end{center}

\vspace*{0.25in}
\noindent {\bf Abstract---}
This is a companion technical report of the main manuscript
``Semidefinite Relaxation and Approximation Analysis of a Beamformed Alamouti Scheme for Relay Beamforming Networks''.
The report serves to give detailed derivations of Lemma 1-2 in the main manuscript, which are too long to be included in the latter.
In addition, more simulation results are presented to verify the viability of the BF Alamouti AF schemes developed in the main manuscript.
\vspace*{0.25in}

\newpage

In the main paper~\cite{WuSOPan2016}, we propose the BF Alamouti AF scheme for the one-hop one-way relay networks. Specifically, the new AF scheme aims at exploring $2$-DoF in the relay AF structure for improving users' SINRs. To analyze the system performance, Theorem 1 is provided in the main paper to prove the SDR approximation bound for the two-variable QCQP problem. Due to the page limit, two important lemmas for proving Theorem 1 is relegated to this companion technical report in Sections \ref{pro:lemma3} and \ref{pro:lemma4}. We also provide some explanations for the system model and supplementary simulation results in Sections \ref{sec1} and \ref{sim}.

\section{Detecting the Alamouti Code Block in the BFA AF Scheme}
In this section, we take the MIMO relay as an example and demonstrate the maximum likelihood detection for the Alamouti code block.  Recall that  with ${\bm V}_p$ defined in the main paper as
${\bm V}_p = [{\bm v}_{1}^p, ...,{\bm v}_{\ell}^p,...{\bm v}_{L}^p]$ and ${\bm v}_{\ell}^p = [v_{1, \ell}^{p},..., v_{\ell, \ell}^{p},...,v_{L, \ell}^{p}]^T$, 
the receive signal at user $(k, i)$ for the MIMO relay case can be expressed as 
\begin{align}\label{chapter2:eq:model_alam_uni_relay_mimo}
{\bf y}_{k, i}(m) \quad =\quad& [~ y_{k, i}(2m), ~ y_{k, i}(2m+1) ~] \\ \notag
\quad=\quad  & \underbrace{\sum_{\ell=1}^L\sum_{c=1}^{L}[({{g}_{k, i}^\ell})^*v_{\ell, c}^1{f}_k^c, ({{g}_{k, i}^\ell})^*v_{\ell, c}^2{{f}_k^c}^*]  \begin{bmatrix}  s_k(2m) &  s_k(2m+1) \\ - s_k(2m+1)^* &  s_k(2m)^* \end{bmatrix}}_{\rm desired~signal} \\ \notag
\quad&+\quad\underbrace{\sum_{\ell=1}^L\sum_{c=1}^{L}\sum_{j \neq k}[({{g}_{k, i}^\ell})^*v_{\ell, c}^1{f}_j^c, ({{g}_{k, i}^\ell})^*v_{\ell, c}^2{{f}_j^c}^*]  \begin{bmatrix}  s_j(2m) &  s_j(2m+1) \\ - s_j(2m+1)^* &  s_j(2m)^* \end{bmatrix}}_{\rm interference~signal} \\ \notag
\quad &+\underbrace{\sum_{\ell=1}^L\sum_{c=1}^{L}[({{g}_{k, i}^\ell})^*v_{\ell, c}^1, ({{g}_{k, i}^\ell})^*v_{\ell, c}^2]  \begin{bmatrix}  {n}^\ell(2m) &  {n}^\ell(2m+1) \\ - {n}^\ell(2m+1)^* &  {n}^\ell(2m)^* \end{bmatrix} + [{{\mu}_{k, i}}(2m), ~ {{\mu}_{k, i}}(2m+1)]}_{\rm noise}.
\end{align}
Denote $h_{1}^j = \sum_{\ell=1}^L\sum_{c=1}^{L}({{g}_{j, i}^\ell})^*v_{\ell, c}^1{f}_j^c$, $h_{2}^j = \sum_{\ell=1}^L\sum_{c=1}^{L}({{g}_{j, i}^\ell})^*v_{\ell, c}^2{{f}_j^c}^*$, $u_{1}^\ell = \sum_{c=1}^{L}({{g}_{k, i}^\ell})^*v_{\ell, c}^1$ and  $u_{2}^\ell = \sum_{c=1}^{L}({{g}_{k, i}^\ell})^*v_{\ell, c}^2$. Then, we can rewrite
\begin{align}\label{chapter2:eq:model_alam_uni_relay_mimo}
{\bf y}_{k, i}(m) \quad =\quad& [~ y_{k, i}(2m), ~ y_{k, i}(2m+1) ~] \\ \notag
\quad=\quad  & [h_{1}^k , h_{2}^k ]  \begin{bmatrix}  s_k(2m) &  s_k(2m+1) \\ - s_k(2m+1)^* &  s_k(2m)^* \end{bmatrix} \\ \notag
\quad&+\quad\sum_{j \neq k}[h_{1}^j , h_{2}^j ]  \begin{bmatrix}  s_j(2m) &  s_j(2m+1) \\ - s_j(2m+1)^* &  s_j(2m)^* \end{bmatrix} \\ \notag
\quad &+\sum_{\ell=1}^L[u_{1}^\ell, u_{2}^\ell]  \begin{bmatrix}  {n}^\ell(2m) &  {n}^\ell(2m+1) \\ - {n}^\ell(2m+1)^* &  {n}^\ell(2m)^* \end{bmatrix} + [{{\mu}_{k, i}}(2m), ~ {{\mu}_{k, i}}(2m+1)].
\end{align}
Denoting ${\bm H}_i = \begin{bmatrix}  h_{1}^i &  -h_{2}^i \\ (h_{2}^i)^* &  (h_{1}^i)^* \end{bmatrix}$ and ${\bm U}_\ell = \begin{bmatrix}  u_{1}^\ell &  -u_{2}^\ell \\ (u_{2}^\ell)^* &  (u_{1}^\ell)^* \end{bmatrix}$, the detection is performed by rewriting the receive signal as
\begin{align*}
\left(
\begin{array}{c}
y_{k, i}(2m)\\
y_{k, i}(2m+1)^* 
\end{array}
\right) = &{\bm H}_i
\left(
\begin{array}{c}
s_k(2m)\\
s_k(2m+1)^* 
\end{array}
\right) + \sum_{j \neq k}{\bm H}_j
\left(
\begin{array}{c}
s_j(2m)\\
s_j(2m+1)^* 
\end{array} 
\right)\\
+&\sum_{\ell=1}^L{\bm U}_\ell
\left(
\begin{array}{c}
{n}^\ell(2m)\\
{n}^\ell(2m+1)^* 
\end{array}
\right) + \left(
\begin{array}{c}
{{\mu}_{k, i}}(2m)\\
{{\mu}_{k, i}}(2m+1)^* 
\end{array}
\right).
\end{align*}
Then, we have
\begin{align*}
&{\bm H}_k^H
\left(
\begin{array}{c}
y_{k, i}(2m)\\
y_{k, i}(2m+1)^* 
\end{array}
\right) = {\bm H}_k^H {\bm H}_k
\left(
\begin{array}{c}
s_k(2m)\\
s_k(2m+1)^* 
\end{array}
\right)+ \sum_{j \neq k}{\bm H}_k^H{\bm H}_j
\left(
\begin{array}{c}
s_j(2m)\\
s_j(2m+1)^* 
\end{array}
\right)\\
&+\sum_{\ell=1}^L{\bm H}_k^H{\bm U}_\ell
\left(
\begin{array}{c}
{n}^\ell(2m)\\
{n}^\ell(2m+1)^* 
\end{array}
\right) + {\bm H}_k^H\left(
\begin{array}{c}
{{\mu}_{k, i}}(2m)\\
{{\mu}_{k, i}}(2m+1)^* 
\end{array}
\right).
\end{align*}
Due to the orthogonality of the Alamouti code, we have ${\bm H}_i^H{\bm H}_i ={\bm H}_i{\bm H}_i^H= (|h_{1}^i|^2 + |h_{1}^i|^2){\bm I}$ and ${\bm U}_\ell^H{\bm U}_\ell={\bm U}_\ell{\bm U}_\ell^H = ( |u_{1}^\ell|^2 + |u_{2}^\ell|^2 ){\bm I}$. Note that, we always consider the scenario $|h_{1}^i|^2 + |h_{1}^i|^2 \neq 0, \forall i$, since otherwise, the problem is trivial. Therefore, we can detect $s_k$ by
\begin{align}
&\left(
\begin{array}{c}
\hat s_k(2m)\\
\hat s_k(2m+1)^* 
\end{array}
\right) = (|h_{1}^k|^2 + |h_{1}^k|^2)^{-1}{\bm H}_k^H
\left(
\begin{array}{c}
y_{k, i}(2m)\\
y_{k, i}(2m+1)^* 
\end{array}
\right) \notag\\\label{hats}
= &
\left(
\begin{array}{c}
s_k(2m)\\
s_k(2m+1)^* 
\end{array}
\right)+ \underbrace{(|h_{1}^k|^2 + |h_{1}^k|^2)^{-1}{\bm H}_k^H\sum_{j \neq k}{\bm H}_j
\left(
\begin{array}{c}
s_j(2m)\\
s_j(2m+1)^* 
\end{array}
\right)}_{\text{interference}}\\\notag
&+\underbrace{\sum_{\ell=1}^L(|h_{1}^k|^2 + |h_{1}^k|^2)^{-1}{\bm H}_k^H{\bm U}_\ell
\left(
\begin{array}{c}
{n}^\ell(2m)\\
{n}^\ell(2m+1)^* 
\end{array}
\right) + (|h_{1}^k|^2 + |h_{1}^k|^2)^{-1}{\bm H}_k^H\left(
\begin{array}{c}
{{\mu}_{k, i}}(2m)\\
{{\mu}_{k, i}}(2m+1)^* 
\end{array}
\right)}_{\text{noise}}.
\end{align}
Since the transmit signals and noise are i.i.d. random variables, in \eqref{hats} we have the signal power is equal to $1$, the interference (treating as noise) power is
\begin{align*}
&(|h_{1}^k|^2 + |h_{1}^k|^2)^{-1}{\bm H}_k^H\sum_{j \neq k}{\bm H}_j {\bm H}_j^H{\bm H}_k (|h_{1}^k|^2 + |h_{1}^k|^2)^{-1} =\frac{\textstyle\sum_{j \neq k}\left(|h_{1}^j|^2 + |h_{1}^j|^2\right)}{|h_{1}^k|^2 + |h_{1}^k|^2}
\end{align*}
and the noise power is given by
\begin{align*}
\sum_{\ell=1}^L\sigma_{\ell}^2(|h_{1}^k|^2 + |h_{1}^k|^2)^{-2}{\bm H}_k^H{\bm U}_\ell{\bm U}_\ell^H{\bm H}_k
+ \sigma_{k, i}^2(|h_{1}^k|^2 + |h_{1}^k|^2)^{-1}=\frac{\textstyle\sum_{\ell=1}^L\sigma_{\ell}^2\left(|u_{1}^\ell|^2 + |u_{2}^\ell|^2\right) + \sigma_{k, i}^2}{|h_{1}^k|^2 + |h_{1}^k|^2}.
\end{align*}
Therefore, the SINR can be expressed as
\[
\frac{|h_{1}^k|^2+|h_{2}^k|^2}{\sum_{j \neq k}(|h_{1}^j|^2+|h_{2}^j|^2) + \sum_{\ell=1}^L\sigma_{\ell}^2 (|u_{1}^\ell|^2 + |u_{2}^\ell|^2) + \sigma_{k, i}^2} = 
\frac{{\bm w}_1^H{\bm A}_{k,i}{\bm w}_1 + {\bm w}_2^H\bar{\bm A}_{k,i}{\bm w}_2}{{\bm w}_1^H{\bm C}_{k,i}{\bm w}_1 + {\bm w}_2^H\bar{\bm C}_{k,i}{\bm w}_2+1},
\]
where ${\bm w}_1 = {\rm vec}({\bm V}_1)$,${\bm w}_2 = {\rm vec}({\bm V}_2)$ and
\begin{align}\notag
{\bm A}_{k,i} &= P_k({\bm f}_k^* \otimes {\bm g}_{k,i})({\bm f}_k^* \otimes {\bm g}_{k,i})^H/\sigma_{k,i}^2 \\ \label{ck_v}
{\bm C}_{k,i} &= \displaystyle \sum_{m \neq k}P_m({\bm f}_m^* \otimes {\bm g}_{k,i})({\bm f}_m^* \otimes {\bm g}_{k,i})^H/\sigma_{k,i}^2 + {\bm \Sigma} \otimes ({\bm g}_{k,i}{\bm g}_{k,i})^H/{\sigma_{k,i}^2} \\\notag
\bar {\bm A}_{k,i} & = P_k({\bm f}_k \otimes {\bm g}_{k,i})({\bm f}_k \otimes {\bm g}_{k,i})^H/\sigma_{k,i}^2 \\\notag
\bar {\bm C}_{k,i} & = \displaystyle \sum_{m \neq k}P_m({\bm f}_m \otimes {\bm g}_{k,i})({\bm f}_m \otimes {\bm g}_{k,i})^H/\sigma_{k,i}^2 + {\bm \Sigma} \otimes ({\bm g}_{k,i}{\bm g}_{k,i})^H)/{\sigma_{v,i}^2}.
\end{align}

\section{Exact Expressions of the Received Signals at User-$(k, i)$ for the Distributed Relay Network}\label{sec1}
To help the readers to well understand the Alamouti BF AF structure for the distributed relay network, we denote
\[
{\bm w}_p= {\rm Diag} ({\bm V}^p), \quad p=1,2,
\]
and
\[
{\bm w}_p = [w_{1}^p, ..., w_{\ell}^p,...,w_{L}^p]^T \in \mathcal{C}^L
\]
as the AF weight at time slot $p$.
Then, we can rewrite the receive signals at user-$(k, i)$ as follows
\begin{align}\label{chapter2:eq:model_alam_uni_relay}
&{\bf y}_{k, i}(m) \quad = [~ y_{k, i}(2m), ~ y_{k, i}(2m+1) ~] \\ \notag
=&\sum_{\ell=1}^L({{g}_{k, i}^\ell})^*[w_{\ell}^1, w_{\ell}^2]\mathbf{C}( {\bf r}_{\ell}(m) ) + [{{v}_{k, i}}(2m), ~ {{v}_{k, i}}(2m+1)]
,\\ \notag
\quad=\quad  &\sum_{\ell=1}^L({{g}_k^\ell})^*[w_{\ell}^1, w_{\ell}^2]  \begin{bmatrix} r^\ell(2m) & r^\ell(2m+1) \\ -r^\ell(2m+1)^* & r^\ell(2m)^* \end{bmatrix} + [{{v}_{k, i}}(2m), ~ {{v}_{k, i}}(2m+1)], \\ \notag
\quad=\quad  &\underbrace{\sum_{\ell=1}^L[({{g}_{k, i}^\ell})^*w_{\ell}^1{f}_k^\ell, ({{g}_{k, i}^\ell})^*w_{\ell}^2{{f}_k^\ell}^*]  \begin{bmatrix}  s_k(2m) &  s_k(2m+1) \\ - s_k(2m+1)^* &  s_k(2m)^* \end{bmatrix}}_{\rm desired~signal} \\ \notag
\quad&+\underbrace{\sum_{\ell=1}^L\sum_{j \neq k}[({{g}_{k, i}^\ell})^*w_{\ell}^1{f}_k^\ell, ({{g}_{k, i}^\ell})^*w_{\ell}^2{{f}_k^\ell}^*]  \begin{bmatrix}  s_j(2m) &  s_j(2m+1) \\ - s_j(2m+1)^* &  s_j(2m)^* \end{bmatrix}}_{\rm interference~signal} \\ \notag
  \quad &+\underbrace{\sum_{\ell=1}^L[({{g}_{k, i}^\ell})^*w_{\ell}^1, ({{g}_{k, i}^\ell})^*w_{\ell}^2]  \begin{bmatrix}  {n}^\ell(2m) &  {n}^\ell(2m+1) \\ - {n}^\ell(2m+1)^* &  {n}^\ell(2m)^* \end{bmatrix} + [{{v}_{k, i}}(2m), ~ {{v}_{k, i}}(2m+1)]}_{\rm noise}。
\end{align}
Note that the SINR expression for the distributed relay case can be derived in a similar way as the MIMO relay in Section 1, which leads to the SINR expression at user-$({k,i})$ as
\begin{align}
\frac{{\bm w}_1^H{\bm A}_{k,i}{\bm w}_1 + {\bm w}_2^H\bar{\bm A}_{k,i}{\bm w}_2}{{\bm w}_1^H{\bm C}_{k,i}{\bm w}_1 + {\bm w}_2^H\bar{\bm C}_{k,i}{\bm w}_2+1}
\end{align}
where 
\begin{align} \notag
{\bm A}_{k,i} = &P_k({\bm f}_k^*\odot {\bm g}_{k,i})({\bm f}_k^*\odot {\bm g}_{k,i})^H/\sigma_{k,i}^2 \\ \notag
{\bm C}_{k,i} = &\sum_{m \neq k}P_m({\bm f}_m^*\odot {\bm g}_{k,i})({\bm f}_m^*\odot {\bm g}_{k,i})^H/\sigma_{k,i}^2 \\ \notag
\bar {\bm A}_{k, i}  = & P_k({\bm f}_k\odot {\bm g}_{k,i})({\bm f}_k\odot {\bm g}_{k,i})^H/\sigma_{k,i}^2 \\\notag
\bar {\bm C}_{k,i}  = & \displaystyle\sum_{m \neq k}P_m({\bm f}_m\odot {\bm g}_{k,i})({\bm f}_m\odot {\bm g}_{k,i})^H/\sigma_{k,i}^2 \\\notag
&+{\rm Diag} (|g_{k, i}^1|^2\sigma_1^2, |g_{k, i}^2|^2\sigma_2^2,...,|g_{k, i}^L|^2\sigma_L^2)/\sigma_{k,i}^2.
\end{align}

\section{Derivation for Theorem 1}

\subsection{Proof of Lemma 1}\label{pro:lemma3}
This proof can be seen as a nontrivial generalization of Theorem 1 in \cite{WuLiMaSo2015SPAWC}. To begin the proof, let $\bar{\bm Q}$, $\tilde{\bm Q}$ be unitary matrices which satisfy 
\[
({\bm X}_1^\star)^{1/2}{\bm A}({\bm X}_1^\star)^{1/2}={\bm Q}_1^H\bar{\bm \Lambda}{\bm Q}_1
\] 
and 
\[
({\bm X}_2^\star)^{1/2}\bar{\bm A}({\bm X}_2^\star)^{1/2}={\bm Q}_2^H\tilde{\bm \Lambda}{\bm Q}_2,
\]
where $\bar{\bm \Lambda}={\rm Diag}(\bar\lambda_1,0,\ldots,0)$ and $\tilde{\bm \Lambda}={\rm Diag}(\tilde\lambda_1, 0,\ldots,0)$ since ${\rm rank}({\bm A})=1$ and ${\rm rank}(\bar{\bm A})=1$.  We may consider ${\bm \xi} \sim ({\bm X}_1^\star)^{1/2}{\bm Q}_1^H{\bf x}$ and ${\bm \eta} \sim ({\bm X}_2^\star)^{1/2}{\bm Q}_2^H{\bf y}$ where ${\bf x},{\bf y} \sim \mathcal{CN}({\bf 0},{\bm I})$ are independent.  Then,  we have
\begin{align*}
&\Pr\left( \frac{{\bm \xi}^H{\bm A}{\bm \xi}+ {\bm \eta}^H\bar{\bm A}{\bm \eta} }{{\bm \xi}^H{\bm C}{\bm \xi} + {\bm \eta}^H\bar{\bm C}{\bm \eta}  +1}\le \frac{\rho \left({\bm A} \bullet {\bm X}_1^\star + \bar{\bm A} \bullet {\bm X}_2^\star\right)}{{\bm C}\bullet{\bm X}_1^\star + \bar{\bm C}\bullet{\bm X}_2^\star +1}\right) \\
=& \Pr\left( \frac{\displaystyle\bar\lambda_1|x_1|^2+\tilde\lambda_1|y_1|^2}{\displaystyle\bar\lambda_1 + \tilde\lambda_2}\le \rho \frac{{\bf x}^H {\bm B}{\bf x}+{\bf y}^H\bar{\bm B}{\bf y}+1}{{\bm B}\bullet {\bm I}+\bar{\bm B}\bullet {\bm I}+1} \right)\triangleq \mathcal{Q},
\end{align*}
where 
$
{\bm B} = {\bm Q}_1({\bm X}_1^\star)^{1/2}{\bm C}({\bm X}_1^\star)^{1/2}{\bm Q}_1^H
$
and 
$
\bar{\bm B} = {\bm Q}_2({\bm X}_2^\star)^{1/2}\bar{\bm C}({\bm X}_2^\star)^{1/2}{\bm Q}_2^H.
$ 
Apparently, we have ${\bm B}\succeq {\bf 0}$ and $\bar{\bm B} \succeq {\bf 0}$. In this way, we may write ${\bm B}=\bar{\bm U}^H\bar{\bm \Sigma}\bar{\bm U}$ and $\bar{\bm B}=\tilde{\bm U}^H\tilde{\bm \Sigma}\tilde{\bm U}$, where $\bar{\bm U}$, $\tilde{\bm U}$ are unitary matrices such that $\bar{\bm \Sigma}={\rm Diag}(\bar\mu_1,\ldots,\bar\mu_{\bar h},0,\ldots,0)$, $\tilde{\bm \Sigma}={\rm Diag}(\tilde\mu_1,\ldots,\tilde\mu_{\tilde h},0,\ldots,0)$ with $\bar\mu_1\ge\cdots\ge\bar\mu_{\bar h}>0$ and $\tilde\mu_1\ge\cdots\ge\tilde\mu_{\tilde h}>0$.  Now, let us define
\begin{align*}
\alpha&=\bar\lambda_1/(\bar\lambda_1+\tilde\lambda_1),\quad \beta=\tilde\lambda_1/(\bar\lambda_1+\tilde\lambda_1),\\
\phi_i&=\bar\mu_i/(\sum_j \bar\mu_j+\sum_j \tilde\mu_j),\quad \forall i=1,...,\bar h,\\
\kappa_i&=\tilde\mu_i/(\sum_j \bar\mu_j+\sum_j \tilde\mu_j),\quad \forall i=1,...,\tilde h.
\end{align*}
Then,
we have 
$$\alpha + \beta =1\quad {\rm and} \quad \sum_{i=1}^{\bar h}\phi_i + \sum_{i=1}^{\tilde h}\kappa_i =1.$$ 
Then, letting ${\bf z}=\bar{\bm U}{\bf x}$, ${\bf s}=\tilde{\bm U}{\bf y}$, we obtain 
\begin{align}\label{min_alpha}
&\mathcal{Q}\\ \notag
\le&\Pr\left( \alpha|x_1|^2+\beta|y_1|^2\le \rho \left(\sum_{i=1}^{\bar h}\phi_i|z_i|^2+\sum_{i=1}^{\tilde h}\kappa_i|s_i|^2 \right) +\rho\right).
\end{align}
In the next, we first consider the case of $\alpha, \beta >0$ and later on we will discuss the case of $\min\{\alpha, \beta\} =0$.
Specifically, let us define $\omega\triangleq \min\{\alpha, \beta\}>0$ and we then have
\begin{align}\notag
&\mathcal{Q}
\le \Pr\left( \omega (|x_1|^2+|y_1|^2)\le \rho \left( \sum_{i=1}^{\bar h}\phi_i\left|\bar U_{i1}x_1+\sum_{j=2}^L\bar U_{ij}x_j\right|^2 \right. \right. \\\notag
&\left. \left. ~~+\sum_{i=1}^{\tilde h}\kappa_i\left|\tilde U_{i1}y_1 +\sum_{j=2}^L\tilde  U_{ij}y_j\right|^2+1\right)\right),
\end{align}
where we define $X_{ij}$ as the $j$th element in the $i$th row of matrix ${\bm X}$. Following the inequality $(a+b)^2 \le 2(a^2+b^2)$, we can obtain \eqref{eq:longeq} on top of the next page.
\begin{align}\label{eq:longeq}
\mathcal{Q}\le& \Pr\left( \omega (|x_1|^2+|y_1|^2)\le 2\rho \left( \sum_{i=1}^{\bar h}\phi_i\left(\left|\bar U_{i1}x_1\right|^2+\left|\sum_{j=2}^L\bar U_{ij}x_j\right|^2\right)+\sum_{i=1}^{\tilde h}\kappa_i\left(\left|\tilde U_{i1}y_1\right|^2+\left|\sum_{j=2}^L\tilde  U_{ij}x_j\right|^2\right)+\frac{1}{2}\right)\right)\\\notag
\le& \Pr\left( \omega (|x_1|^2+|y_1|^2)\le 2\rho \left( \sum_{i=1}^{\bar h}\phi_i\left(|x_1|^2+\left|\sum_{j=2}^L\bar U_{ij}x_j\right|^2\right)+\sum_{i=1}^{\tilde h}\kappa_i\left(|y_1|^2+\left|\sum_{j=2}^L\tilde  U_{ij}x_j\right|^2\right)+\frac{1}{2}\right)\right)\\\notag
=& \Pr\left( \omega (|x_1|^2+|y_1|^2)\le 2\rho \left(|x_1|^2+|y_1|^2 + \sum_{i=1}^{\bar h}\phi_i\left|\sum_{j=2}^L\bar U_{ij}x_j\right|^2+\sum_{i=1}^{\tilde h}\kappa_i\left|\sum_{j=2}^L\tilde  U_{ij}x_j\right|^2+\frac{1}{2}\right)\right)\\\notag
=& \Pr\left( |x_1|^2+|y_1|^2\le \frac{2\rho}{\omega-2\rho} \left(\sum_{i=1}^{\bar h}\phi_i|\sum_{j=2}^L\bar U_{ij}x_j|^2+\sum_{i=1}^{\tilde h}\kappa_i|\sum_{j=2}^L\tilde  U_{ij}x_j|^2+\frac{1}{2}\right)\right).
\end{align}
For standard complex Gaussian variables $x$ and $y$, we have 
\[
\Pr(|x|^2+|y|^2\le t)=1-(t+1)e^{-t}\le t^2/2, \quad \forall t >0.
\] 
Then we can obtain \eqref{Gamma} on top of the next page.
\begin{align}\label{Gamma}
\mathcal{Q}\le&\frac{2\rho^2}{(\omega-2\rho)^2} {\mathbb E}\left[ \left(\sum_{i=1}^{\bar h}\phi_i\left|\sum_{j=2}^L\bar U_{ij}x_j\right|^2+\sum_{i=1}^{\tilde h}\kappa_i\left|\sum_{j=2}^L\tilde  U_{ij}x_j\right|^2+\frac{1}{2}\right)\right]^2.
\end{align}
Now, let us define
\[
W_i = \left| \sum_{j=2}^L \bar U_{ij}x_j \right|^2, \quad W_i' = \left| \sum_{j=2}^L \tilde U_{ij}y_j \right|^2
\]
and compute
\begin{align*}
&{\mathbb E}\left[\left(\sum_{i=1}^{\bar h}\phi_i W_i +\sum_{i=1}^{\tilde h}\kappa_i W_i' +\frac{1}{2}\right)\right]^2 \\
=& {\mathbb E} \left[ \sum_{i=1}^{\bar h} \phi_i W_i \right]^2 + {\mathbb E}\left[ \sum_{i=1}^{\tilde h} \kappa_i W_i' \right]^2 + \frac{1}{4} \\
+&  {\mathbb E}\left[\sum_{i=1}^{\bar h} \phi_i W_i + \sum_{i=1}^{\tilde h} \kappa_i W_i' \right] \\
+&  2{\mathbb E}\left[ \left( \sum_{i=1}^{\bar h} \phi_i W_i  \right)\left(\sum_{i=1}^{\tilde h} \kappa_i W_i' \right) \right].
\end{align*}
It follows that
\begin{align*}
&{\mathbb E} \left[ \sum_{i=1}^{\bar h} \phi_i W_i \right]^2 = {\mathbb E} \left[ \sum_{i,j=1}^{\bar h} \phi_i\phi_j W_iW_j \right]\\
=& \sum_{i,j=1}^{\bar h} \phi_i\phi_j \sum_{k,k'=2}^L \sum_{l,l'=2}^L \bar U_{ik}\bar U_{ik'}^*\bar U_{jl}\bar U_{jl'}^* {\mathbb E}[x_kx_{k'}^*x_lx_{l'}^*] \\
=& \sum_{i,j=1}^{\bar h} \phi_i\phi_j \left(2\sum_{k=2}^L |\bar U_{ik}|^2|\bar U_{jk}|^2 + \sum_{2\le k\not=l \le L} |\bar U_{ik}|^2|\bar U_{jl}|^2 \right) \\
=& \sum_{i,j=1}^{\bar h} \phi_i\phi_j \left(\sum_{k=2}^L |\bar U_{ik}|^2|\bar U_{jk}|^2 + \sum_{k,l=2}^L |\bar U_{ik}|^2|\bar U_{jl}|^2\right),
\end{align*}
where the third equality is derived by calculating ${\mathbb E}[x_kx_{k'}^*x_lx_{l'}^*]$ where $x_k, x_{k'}, x_l, x_{l'}$ are standard complex Gaussian variables.
Similarly, we have
\begin{align*}
&{\mathbb E} \left[ \sum_{i=1}^{\tilde h} \kappa_i W_i \right]^2\\
=&\sum_{i,j=1}^{\tilde h} \kappa_i\kappa_j \left(\sum_{k=2}^L |\tilde U_{ik}|^2|\tilde U_{jk}|^2 + \sum_{k,l=2}^L |\tilde U_{ik}|^2|\tilde U_{jl}|^2\right).
\end{align*}
Therefore, we have
\[
{\mathbb E} \left[ \sum_{i=1}^{\bar h} \phi_i W_i \right]^2 + {\mathbb E} \left[ \sum_{i=1}^{\tilde h} \kappa_i W_i \right]^2 \le 2.
\]
Similarly, we compute
\begin{align*}
&{\mathbb E}\left[ \sum_{i=1}^{\bar h}\phi_i W_i \right] + {\mathbb E}\left[ \sum_{i=1}^{\tilde h}\kappa_i W_i' \right] \\
=& \sum_{i=1}^{\bar h} \phi_i {\mathbb E}\left[ \left| \sum_{j=2}^L \bar U_{ij}x_j \right|^2 \right] +  \sum_{i=1}^{\tilde h} \kappa_i {\mathbb E}\left[ \left| \sum_{j=2}^L \tilde U_{ij}y_j \right|^2 \right] \\
=& \sum_{i=1}^{\bar h} \phi_i \left( \sum_{j=2}^L\sum_{k=2}^L \bar U_{ij}\bar U_{ik}^* {\mathbb E}[x_jx_k^*] \right) \\
&+ \sum_{i=1}^{\tilde h} \kappa_i \left( \sum_{j,k=2}^L \tilde U_{ij}\tilde U_{ik}^* {\mathbb E}[y_jy_k^*] \right) \\
=& \sum_{i=1}^{\bar h} \sum_{j=2}^L \phi_i | \bar U_{ij}|^2 {\mathbb E}[|x_j|^2] + \sum_{i=1}^{\tilde  h} \sum_{j=2}^L \kappa_i | \tilde  U_{ij}|^2 {\mathbb E}[|y_j|^2]\\
\le& 1,
\end{align*}
from which we have
$$ {\mathbb E}\left[ \left( \sum_{i=1}^{\bar h}\phi_i W_i \right)\left( \sum_{i=1}^{\tilde h}\kappa_i W_i' \right)\right]  \le 1, $$
since $\phi_i, \kappa_i, W_i, W_i'\ge 0$. Therefore, we have
\begin{align}\label{rb2}
\mathcal{Q} \le \frac{21\rho^2}{2(\omega-2\rho)^2} \le \left(\frac{4\rho}{\omega-2\rho}\right)^2.
\end{align}
Recall that we have defined $\omega \triangleq \min\{\alpha, \beta\}$ and $\alpha + \beta = 1$. Hence, we have 
\begin{align*}
\omega = \frac{\min\{{\bm A}\bullet{\bm X}_1^\star, \bar{\bm A}\bullet{\bm X}_2^\star\}}{{\bm A}\bullet{\bm X}_1^\star + \bar{\bm A}\bullet{\bm X}_2^\star}
\end{align*} 
as desired.

On the other hand, following \eqref{min_alpha}, we can otherwise bound $\mathcal{Q}$ as
\begin{align}\label{min_alpha1}
&\mathcal{Q}\\ \notag
\le&\Pr\left( |x_1|^2\le 2\rho \left(\sum_{i=1}^{\bar h}\phi_i|z_i|^2+\sum_{i=1}^{\tilde h}\kappa_i|s_i|^2  +1\right)\right).
\end{align} 
Herein, we assume that $\alpha \ge 0.5$ and $\beta \le 0.5$. Note that for a standard complex Gaussian variables $x$, we have 
\[
\Pr(|x|^2\le t)=1-e^{-t} \le t, \quad \forall t >0.
\] 
Then, following Lemma 2 in \cite{chang2008approximation} and previous derivations, we have
\begin{align}\label{Gamma1}
\mathcal{Q}\le&\frac{2\rho}{(1-2\rho)} \left( 1 + 1\right) = \frac{4\rho}{1-2\rho}.
\end{align}

Remember that we still need to consider the case of $\min\{\alpha, \beta\} =0$. Without loss of generality, we assume that $\beta=0$. Then it follows that 
\begin{align}\notag
\mathcal{Q}\le \Pr\left( |x_1|^2\le \rho \left(\sum_{i=1}^{\bar h}\phi_i|z_i|^2+\sum_{i=1}^{\tilde h}\kappa_i|s_i|^2\right)+\rho \right),
\end{align}
which gives rise to
\begin{align}\label{Gamma2}
\mathcal{Q}\le&\frac{\rho}{(1-\rho)} \left( 1 + 1\right) = \frac{2\rho}{1-\rho} < \frac{4\rho}{1-2\rho}.
\end{align}
It is easy to see that when $\min\{\alpha, \beta\} =0$, the bound in \eqref{rb2} reduces to \eqref{Gamma2}. Then, we can combine \eqref{Gamma1} and \eqref{Gamma2} to arrive at the desired result in Lemma 1. \hfill $\blacksquare$

\subsection{Proof of Lemma 2}\label{pro:lemma4}
To proceed this proof, we define $\bar{\bm P}$ and $\tilde{\bm P}$ to be unitary matrices satisfying $({\bm X}_1^\star)^{1/2}{\bm D}({\bm X}_2^\star)^{1/2}=\bar{\bm P}^H\bar{\bm \Delta}\bar{\bm P}$ and $({\bm X}_2^\star)^{1/2}\bar{\bm D}({\bm X}_2^\star)^{1/2}=\tilde{\bm P}^H\tilde{\bm \Delta}\tilde{\bm P}$, where $\bar{\bm \Delta}={\rm Diag}(\bar\delta_1,\ldots, \bar \delta_{\bar q},0,\ldots,0)$, $\bar\delta_1\ge \cdots\ge\bar\delta_{\bar q} > 0$ and $\tilde{\bm \Delta}={\rm Diag}(\tilde\delta_1,\ldots, \tilde\delta_{\tilde q},0,\ldots,0)$, $\tilde\delta_1\ge \cdots\ge\tilde\delta_{\tilde q} > 0$. Then, we may consider ${\bm \xi} \sim ({\bm X}_1^\star)^{1/2}\bar{\bm P}^H{\bf x}$ and ${\bm \eta} \sim ({\bm X}_2^\star)^{1/2} \tilde{\bm P}^H{\bf y}$ with ${\bf x},{\bf y} \sim \mathcal{CN}({\bf 0},{\bm I})$. It follows that
\begin{eqnarray*}
&~&\Pr\left( {\bm \xi}^H{\bm D}{\bm \xi}+{\bm \eta}^H \bar{\bm D} {\bm \eta} \ge v ({\bm D} \bullet {\bm X}_1^\star + \bar{\bm D} \bullet {\bm X}_2^\star  ) \right) \\
&=&\Pr\left( \sum_{i=1}^{\bar q} \bar\delta_i|x_i|^2 + \sum_{i=1}^{\tilde q} \tilde\delta_i|y_i|^2  \ge v (\sum_{i=1}^{\bar q} \bar\delta_i +  \sum_{i=1}^{\tilde q} \tilde\delta_i ) \right).
\end{eqnarray*}
Note that herein $z_i$ and $s_i$ are standard complex Gaussian variables. 
Let $\alpha_i = \bar\delta_i/(\sum_{j=1}^{\bar q} \bar\delta_j+\sum_{j=1}^{\tilde q} \tilde\delta_j)$ and $\beta_i = \tilde\delta_i/(\sum_{j=1}^{\bar q} \bar\delta_j+\sum_{j=1}^{\tilde q} \tilde\delta_j)$. We have $\sum_{i} \alpha_i + \sum_{i}\beta_i=1$ and
\begin{eqnarray*}
&~&\Pr\left( {\bm \xi}^H{\bm D} {\bm \xi} + {\bm \eta}^H\bar{\bm D} {\bm \eta}  \ge v ({\bm D} \bullet {\bm X}_1^\star + \bar{\bm D} \bullet {\bm X}_2^\star  ) \right) \\
&=&\Pr\left( \sum_{i=1}^{\bar q} \alpha_i|x_i|^2 + \sum_{i=1}^{\tilde q} \beta_i|y_i|^2  \ge v \right)\\
&=&\Pr\left( \sum_{i=1}^{\bar q + \tilde q} \sum_{k=1}^2\nu_i|\epsilon_k|^2  \ge v \right),
\end{eqnarray*}
where $\epsilon_k \sim {\mathcal{N}(0, \frac{1}{2})}$.
Then, using the argument in the proof
of \cite[Proposition
2.1]{so2008unified}) (see the remark after the proof of \cite[Proposition
2.2]{so2008unified}) and the proof of Theorem 1 in \cite{wu2012rank}, we see that for $v \ge 2$, we have the desired result in Lemma 2.
\hfill $\blacksquare$

\section{Further Simulation Results}\label{sim}
In this section, we provide numerical simulations to compare the performance of different AF schemes and demonstrate the superiority of the proposed BF Alamouti AF scheme. Specifically,  some numerical results for the distributed relay network has been provided in the main manuscript and here we show the numerical results for the MIMO relay network and some missing simulation results for distributed relay network.  We assume w.l.o.g. that each multicast group has an equal number of users (i.e., $m_k=M/G$ for $k=1,\ldots,G$).  The channels ${\bm f}_k, {\bm g}_{k,i}$, where $k=1,\ldots,G$ and $i=1,\ldots,m_k$, are identical independently distributed (i.i.d.) according to $\mathcal{CN}({\bm 0}, {\bm I})$.  The transmitted signal at each transmitter is with power $0$dB (i.e., $P_j=0$dB for $j=1,\ldots,G$). Each single-antenna relay has the same noise power (i.e., $\sigma_{\ell}^2=\sigma_{\sf ant}^2$, where $\ell=1,\ldots,L$), and all users have the same noise power (i.e., $\sigma_{k,i}^2=\sigma_{\sf user}^2$ for $k=1,\ldots,G$ and $i=1,\ldots,m_k$). We assume that $\sigma_{\sf ant}^2>0$ and $\sigma_{\sf user}^2>0$.  The total power threshold for all the relays is $\bar{P}_0$; the power threshold at $\ell$th relay is $\bar{P}_\ell$, where $\ell=1,\ldots,L$. For each AF scheme, $100$ channel realizations were averaged to get the plots, and the number of randomizations for generating BF AF weights and BF Alamouti weights is $1,000$.

\subsection{Worst User's SINR versus Total Power Threshold} \label{subsec:tot-pow}
\begin{figure}[htb]
\centering
\includegraphics[width=0.6\textwidth]{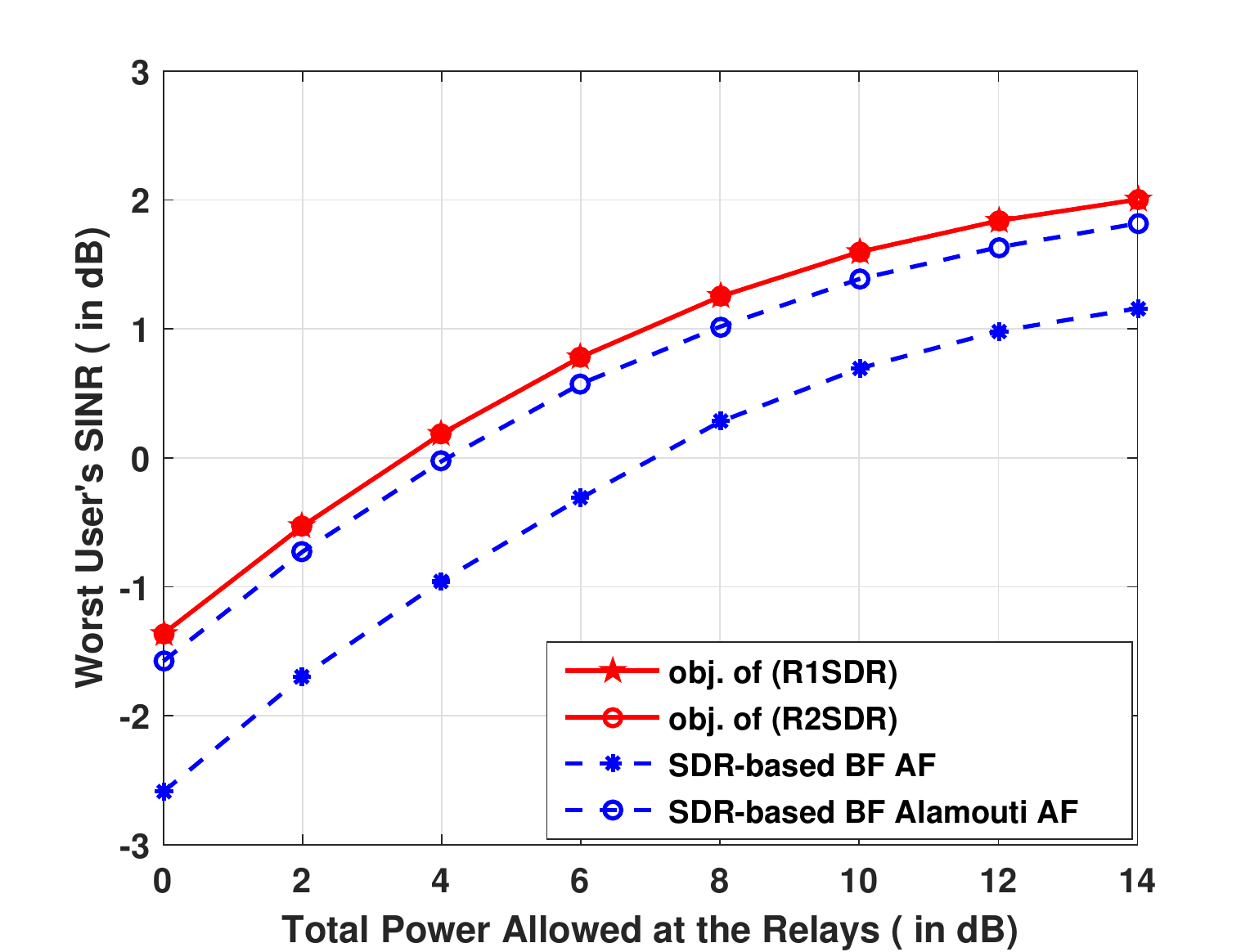}
\caption{Worst user's SINR versus total power threshold at the MIMO relay: $L=4$, $G=2$, $M=16$, $\sigma_{\sf ant}^2=\sigma_{\sf user}^2=0.25$.}
\label{fig:1}
\end{figure}

In this simulation, we vary the total power budget at relays to see the worst user's SINR performance in different relay networks. For ease of exposition, 
we consider the scenario where only the total power constraint is present. For the MIMO relay case in Figure \ref{fig:1}, we assume that there are $L=4$ single-antenna relays and $G=2$ multicast groups with a total of $M=12$ users, i.e., each multicast group has $6$ users. We set $\sigma_{\sf ant}^2=\sigma_{\sf user}^2=0.25$. From the figure,  we see that 
the objective (obj.) of (R1SDR) and (R2SDR) serve as an upper bound for the SDR-based BF AF scheme and the SDR-based BF Alamouti AF scheme, respectively. Moreover, based on randomization, the BF Alamouti AF scheme shows a significant SINR improvement over the BF AF scheme in all the power regions. 

\subsection{Worst User's SINR versus Number of Per-relay Power Constraints}
\begin{figure}[htb]
\centering
\includegraphics[width=0.6\textwidth]{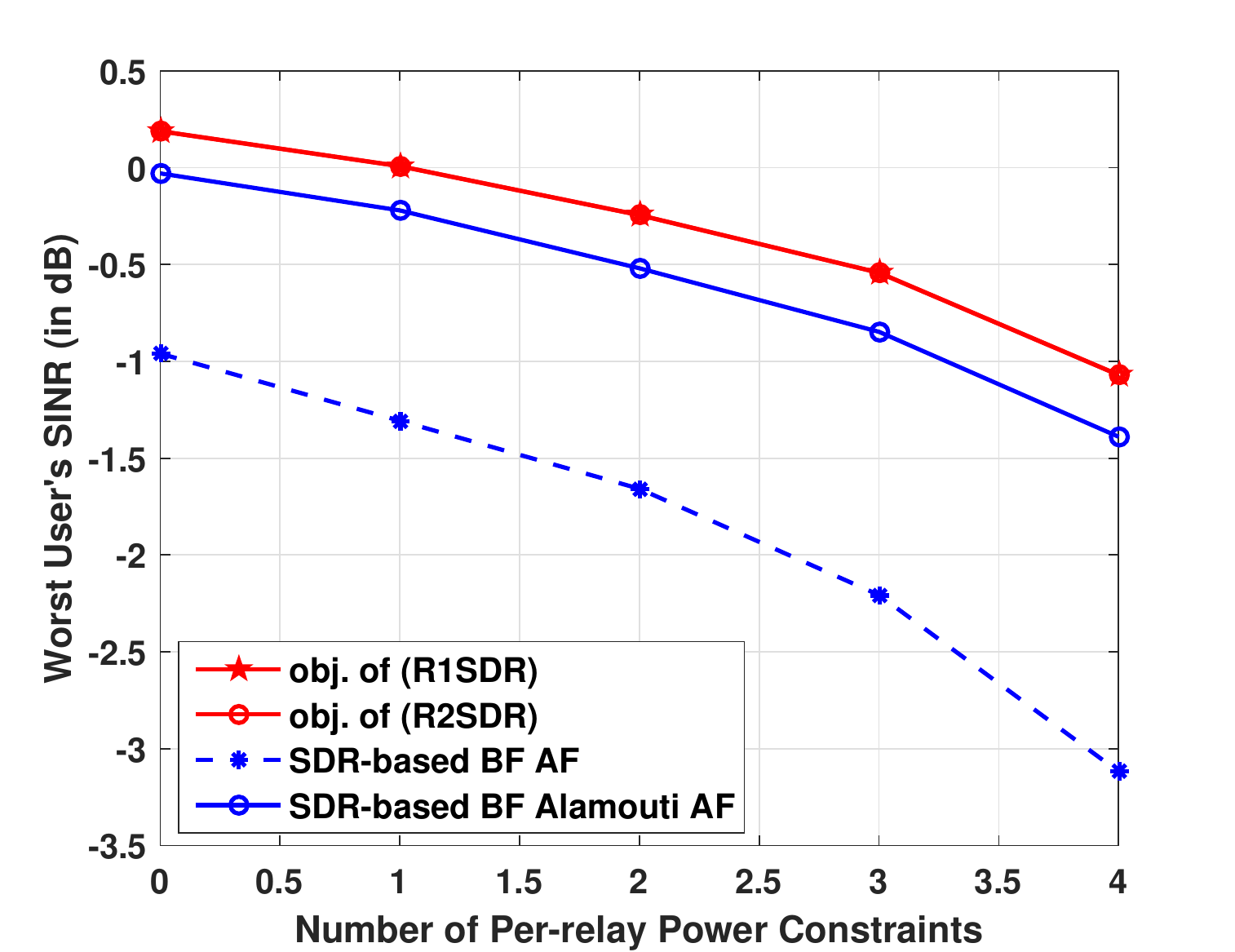}
\caption{Worst user's SINR versus number of per-relay power constraints in the MIMO relay network: $L=4$, $G=2$, $M=16$, $\bar{P}_0=4$dB, $\bar{P}_\ell = -5$dB for $\ell=1,\ldots,L$, $\sigma_{\sf ant}^2=\sigma_{\sf user}^2=0.25$.}
\label{fig:5}
\end{figure}

In this simulation, we consider the scenario where both the total power constraint and per-antenna power constraints are present and the primal users are absent.  Our purpose is to see how the worst user's SINR scales with the number of per-relay power constraints. Specifically, Figure \ref{fig:5} shows the MIMO relay case with $L=4$, $G=2$, $M=16$, where the total power threshold is $\bar{P}_0=4$dB and the per-relay power threshold is $-5$dB for all relays (i.e., $\bar{P}_1=\cdots=\bar{P}_L=-5$dB).  We set $\sigma_{\sf ant}^2=\sigma_{\sf user}^2=0.25$ and vary the number of per-relay power constraints from $0$ to $L$ to compare SINR performances of different AF schemes.  It shows that the BF Alamouti AF scheme outperforms the BF AF scheme.  As the number of per-relay power constraints increases, the SINRs diverge from their SDR upper bounds, and both BF AF and BF Alamouti AF exhibit the same scaling with $L$, which is consistent with the approximation bounds in terms of $J$ in Proposition 1 and Theorem 1 in the main paper.

\subsection{Worst User's SINR versus Number of Primal Users}
\begin{figure}[htb]
\centering
\includegraphics[width=0.6\textwidth]{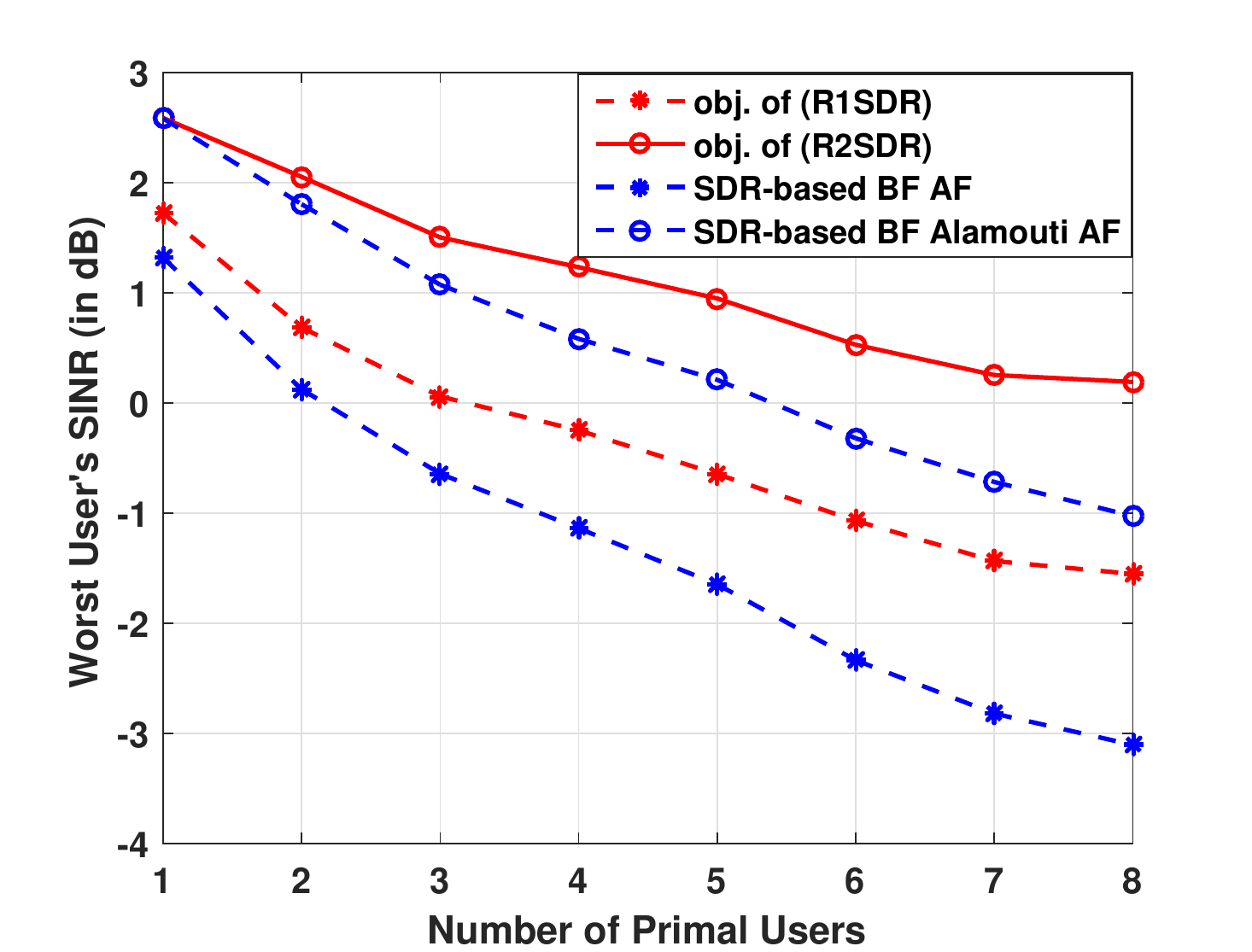}
\caption{Worst user's SINR versus number of primal users in the MIMO relay CR network: $L=4$, $G=2$, $M=12$, $\bar{P}_0=10$dB, $b_u = 3$dB for $u=1,\ldots,U$, $\sigma_{\sf ant}^2=\sigma_{\sf user}^2=0.25$ and $\sigma_{\sf u}^2=0.25$.}
\label{fig:7}
\end{figure}

Similar to previous simulations, here we show the worst user's SINR scaling with the number of primal users. To set up the problem, we consider the scenario where the total power constraint and the primal users' interference constraints are present.  We assume that $L=4$, $G=2$ and $M=12$ in the MIMO relay network. We set $\sigma_{\sf ant}^2=\sigma_{\sf user}^2=0.25$,  the total power budget $\bar{P}_0=10$dB, and the noise power at all primal users to be $\sigma_{\sf u}^2=0.25$. Moreover, we assume that the primal users are subject to an interference power threshold equaling to $b_u=3$dB. We then increase the number of primal users to see the SINR scaling in Figure~\ref{fig:7}. It shows that as the primal users increases, both the BF AF scheme and BF Alamouti AF scheme diverge from their SDR bounds and BF Alamouti AF shows a significant improvement over BF AF. These results further validate Proposition 1 and Theorem 1 in terms of the scaling of $J$.

\subsection{Actual Bit Error Rate (BER) Performance} \label{subsec:BER}
\begin{figure}[htb]
\centering
\includegraphics[width=0.6\textwidth]{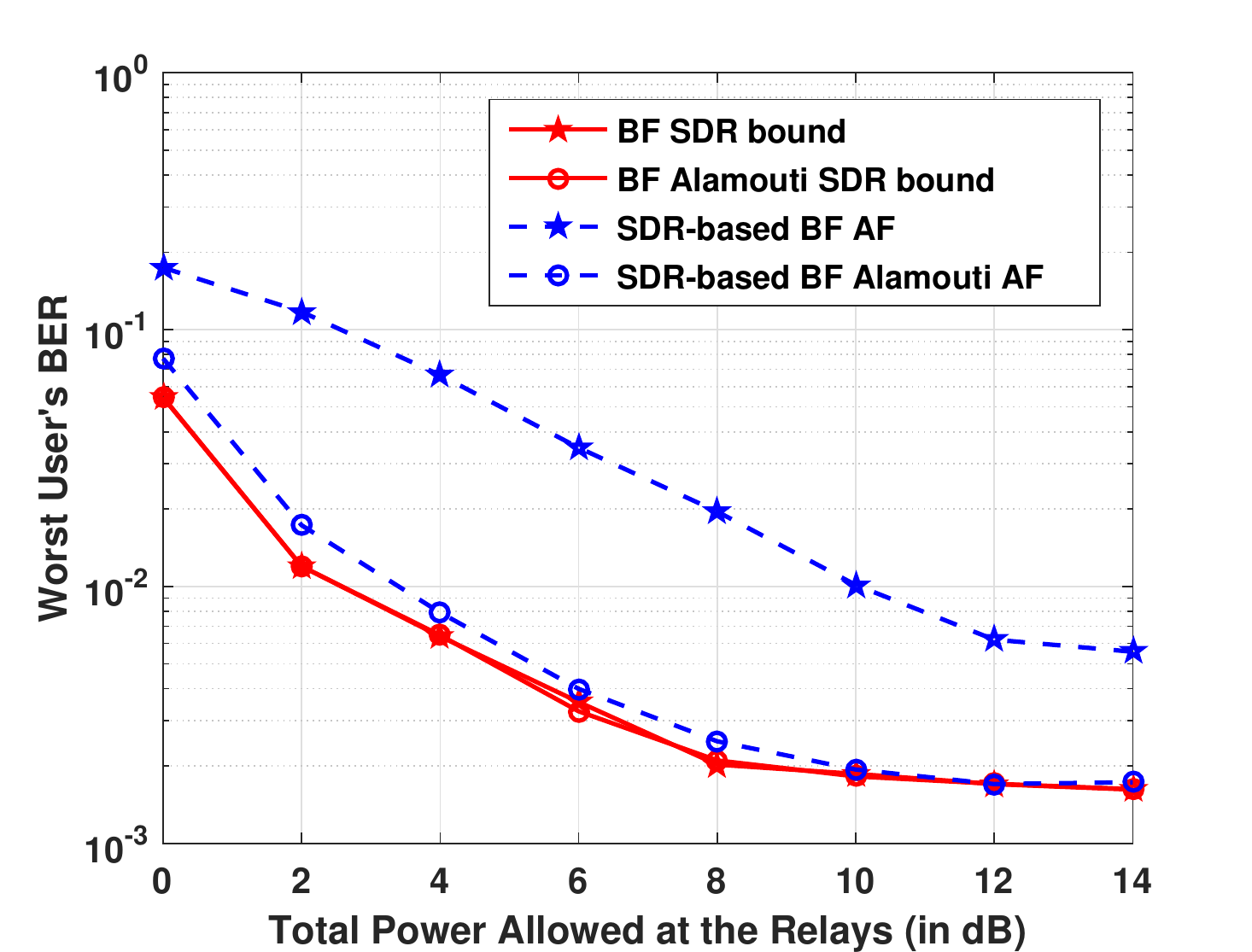}
\caption{Worst user's BER achieved by different AF schemes versus total power threshold at the MIMO relay: $L=4$, $G=2$, $M=16$, $\sigma_{\sf ant}^2=\sigma_{\sf user}^2=0.25$. A rate-$\frac{1}{3}$ turbo code with codelength $2880$ is used.}
\label{fig:13}
\end{figure}

To further demonstrate the efficacy of the proposed AF scheme, we study the actual coded bit error rate (BER) performance of the scenario setting in Figures \ref{fig:1}. The resulting BERs are shown in Figure~\ref{fig:13}.  To simulate the SDR bound in the BER plots, we assume that there exists an SISO channel whose SINR is equal to $\gamma({\bm W}^\star)$ or $\theta({\bm W}_1^\star, {\bm W}_2^\star)$. In our simulations, we adopt a gray-coded QPSK modulation scheme and a rate-$1/3$ turbo code in \cite{Std:16e} with a codelength of $2880$ bits.  We simulate $100$ code blocks for each channel realization and thus the BER reliability level is $10$e$-4$. We see that the actual BER performance of the proposed BF Alamouti AF scheme indeed outperforms the BF-AF scheme at almost all power thresholds.  The results are consistent with those SINR results in Figure~\ref{fig:1} and show that BF Alamouti AF can achieve a good performance in real applications.

\subsection{A Comparison with the Feasible Point Pursuit (FPP) Algorithm} \label{subsec:FPP}
\begin{figure}[htb]
\centering
\includegraphics[width = 0.6\textwidth]{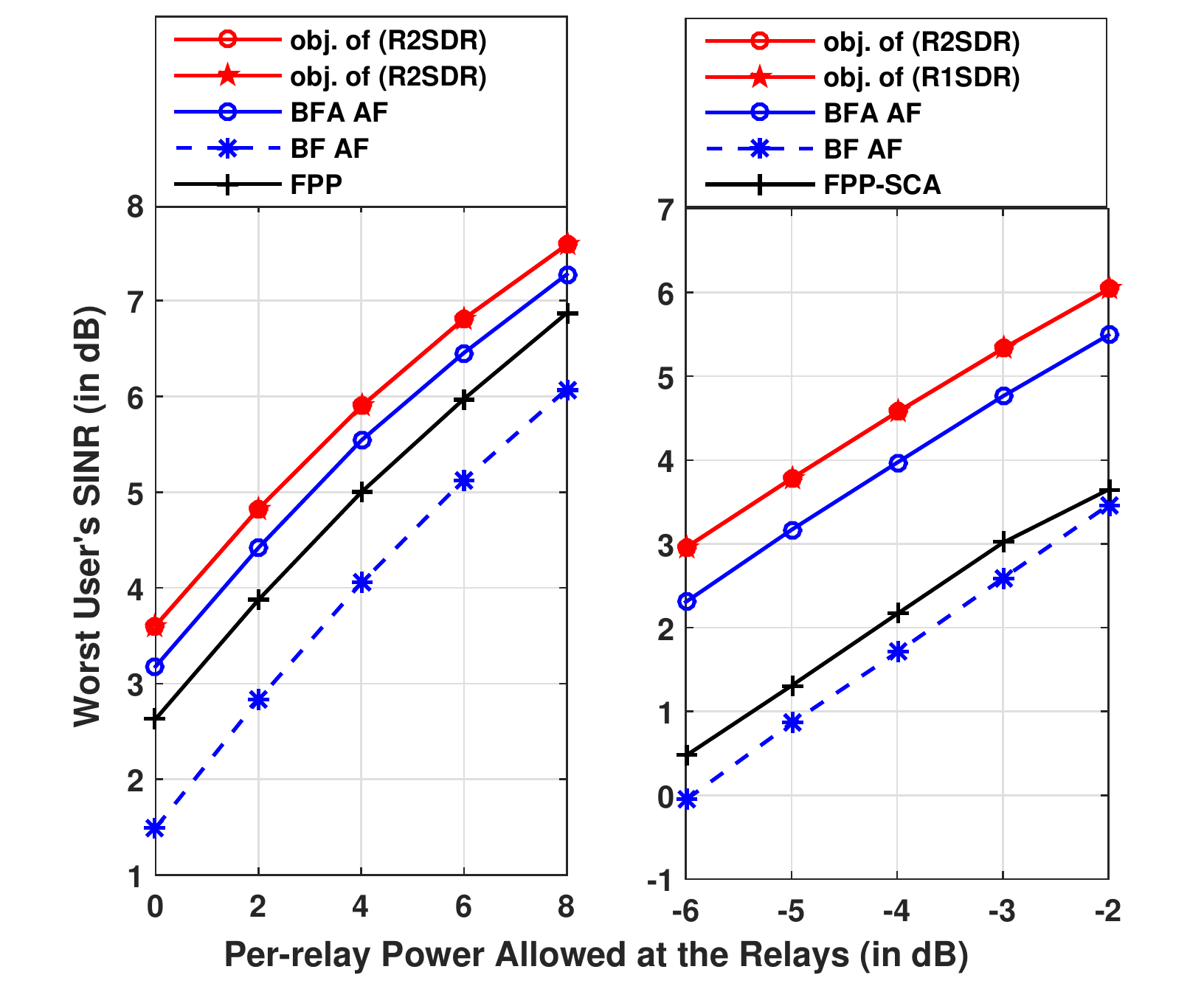}
\caption{Comparison with the feasible point pursuit method.}
\label{fig:fpp}
\end{figure}

In this paper, we compare the proposed BF Alamouti AF scheme with the art-of-the-art algorithm for solving one-variable QCQPs.
Specifically, we show the comparison results of the BF Alamouti AF scheme with the FPP scheme in \cite{mehanna2015feasible} in a distributed relay network and with the FPP-SCA scheme in \cite{christopoulos2015multicast} in an MIMO relay network. In the left sub-figure of Figure \ref{fig:fpp}, we consider only the total power constraint and use the system setting $L=8$, $G=1$, $M=16$, $\sigma_{\sf ant}^2=\sigma_{\sf user}^2=0.25$. In the right sub-figure of Figure \ref{fig:fpp},  the system setting is $L=4$, $G=1$, $M=16$, $\sigma_{\sf ant}^2=\sigma_{\sf user}^2=0.25$, $\bar{P}_0=3$dB, and we consider both the total power constraint and per-relay power constraints. Note that the per-relay power thresholds are the same for all relays (i.e., $\bar{P}_1=\cdots=\bar{P}_L$). The results show that the BF Alamouti AF scheme exhibits a big performance gain over the FPP schemes.  We remark here that the FPP schemes has been numerically proven to outperform most of the existing algorithms for solving one-variable QCQPs.

\end{document}